\documentclass[reprint,twocolumn,aps,prb,superscriptaddress]{revtex4-2}
\usepackage[T1]{fontenc}
\usepackage[utf8]{inputenc}
\usepackage{lmodern}
\usepackage{graphicx}
\usepackage{amsmath}
\usepackage{xcolor}
\usepackage{float}
\usepackage{hyperref}
\usepackage{braket}
\usepackage{amssymb,amsmath,amsthm, array}
\usepackage{mdframed}
\usepackage{systeme,mathtools}
\usepackage{yfonts}
\usepackage{comment}

\usepackage{braket}

\def  \bnabla   {\mbox{\boldmath$\nabla $}}

\usepackage{soul}

\DeclareUnicodeCharacter{0301}{-}

\begin{document}
\renewcommand{\vec}[1]{\mathbf{#1}}
\newcommand{\ii}{\mathrm{i}}
\def\ya#1{{\color{orange}{#1}}}

\def\d{\downarrow}
\def\u{\uparrow}
\def\nn{\nonumber}

\title{Quantum Skyrmion Qudit in a Triangular-lattice magnet}

\author{D. Maroulakos}
\affiliation{Doctoral School at the University of Rzeszów, Rzeszów 35-310, Poland}
\author{A. Wal}
\affiliation{Institute of Physics, Faculty of Exact and Technical Sciences, University of Rzeszów, Pigonia 1, Rzeszów 35-310, Poland}
\author{A. Ugulava}
\affiliation{I. Javakhishvili Tbilisi State University, I.Chavchavadze Ave. 3, Tbilisi, 0179, Georgia}
\author{O. Kharshiladze}
\affiliation{I. Javakhishvili Tbilisi State University, I.Chavchavadze Ave. 3, Tbilisi, 0179, Georgia}
\author{L. Chotorlishvili}
\affiliation{Department of Physics and Medical Engineering, Rzesz\'ow University of Technology, 35-959 Rzesz\'ow, Poland}

\date{\today}
\begin{abstract}
Since the pioneering work Lohani \textit{et. al.} Phys. Rev. X \textbf{9}, 041063 (2019), it became clear that quantum skyrmions have highly unusual properties as compared to the classical skyrmions and, due to their quantumness, cannot be described by continuous magnetic textures akin to the classical skyrmions. Competing nearest-neighbor and next-nearest-neighbor ferromagnetic and antiferromagnetic interactions in triangular spin-frustrated magnets lead to the formation of quantum skyrmion states. In frustrated magnets, skyrmions are characterized by the helical degree of freedom, which can store quantum information. In the limit of a weak electric field, the system can be described as a two-level system, i.e., a skyrmion qubit.  Here, we propose a more general formulation of the problem and obtain general analytic solution of the model previously introduced in Psaroudaki \textit{et. al}. Phys. Rev. Lett. \textbf{127}, 067201 (2021). Our solution is valid not only for small barrier but for the arbitrary electric field. In the case of a significant barrier, we prove that the system's state is not a Skyrmion qubit as it was thought before, but a Skyrmion qudit. We constructed the density matrix of the Skyrmion qudit and studied its evolution in time. The obtained results suggest that the proposed model can be exploited further to meet the needs of quantum information theory and quantum skyrmionics. We showed that the $l_1$ norm of coherence of the skyrmion quantum qudit is a thousand times larger than the coherence of the skyrmion quantum qubit. The obtained result opens new perspectives for quantum skyrmion-based resource theory.

\end{abstract}

\maketitle

\section{Introduction}

Skyrmionics is a field dedicated to the study of magnetic topological solitons \cite{PhysRevB.97.064403}. Emerged during the last decade, beyond fundamental aspects it promises plethora of futuristic platforms and applications, such as environmentally friendly nanodevices and the energy harvesting technologies \cite{seki2012observation,PhysRevB.89.094411,PhysRevB.90.094423,leonov2015multiply,PhysRevX.7.041045,PhysRevB.87.024402,muhlbauer2009skyrmion,PhysRevLett.125.227201,PhysRevLett.129.126101}.
Interest in quantum skyrmionics appeared relatively recently \cite{PhysRevX.9.041063,PhysRevB.103.L060404, PhysRevResearch.4.023111, PhysRevB.107.L100419,PhysRevLett.133.043601,PhysRevLett.134.186701,PhysRevLett.133.216702,PhysRevLett.129.017201,PhysRevB.111.134410,PhysRevResearch.6.023067,PhysRevB.110.104411,PhysRevResearch.4.043113,PhysRevLett.127.067201,kurumaji2019skyrmion}. In frustrated magnets, skyrmions are characterized by the helical degree of freedom, which can store quantum information. Our main interest here concerns the application of quantum skyrmions to the quantum information theory \cite{PhysRevLett.127.067201,PhysRevLett.130.106701}. There are different frustrated triangular-lattice magnets hosting skyrmions, and we refer also to those materials \cite{PhysRevMaterials.5.054401}: Cr/MoS$_2$; Mn/MoS$_2$; Fe/MoS$_2$; Co/MoS$_2$; Fe/WSe$_2$; Mn/WS$_2$. Before proceeding further for the sake of a broad audience, we briefly review the basic facts about skyrmion qubits, and for more details, we refer to \cite{PhysRevLett.127.067201}:
The free energy of the Heisenberg model hosting antiferromagnetic skyrmions has the form \cite{PhysRevLett.127.067201}: 
\begin{eqnarray}\label{1antiferromagnetic skyrmions}
&&F=-\frac{J_1}{2}\left(\bnabla \textbf{m}\right)^2+\frac{J_2a^2}{2}\left(\bnabla^2 \textbf{m}\right)^2-\nonumber\\
&&(H/a^2)m_z+(K/a^2)m_z^2.
\end{eqnarray}
Here, $H$ and $K$ correspond to the Zeeman and anisotropy terms. The competing ferromagnetic and antiferromagnetic exchange interactions ($J_1$ and $J_2$ terms respectively) lead to the formation of non-colinear magnetic texture $\textbf{m}(\textbf{r})$ and the lattice constant hereafter we set to $a=1$ and dimensionless $\textbf{r}\equiv \textbf{r}/(a\sqrt{J_2/J_1})$. The prototype material for Eq.(\ref{1antiferromagnetic skyrmions}) is the frustrated triangular-lattice magnet e.g., $\text{Gd}_2\text{2PdSi}_3$ \cite{kurumaji2019skyrmion}. 
 The typical values of parameters mentioned in the Free energy Eq.(\ref{1antiferromagnetic skyrmions}) are \cite{PhysRevLett.127.067201}: $J_{1,2}=1$meV, $a=5\text{\AA}$, the skyrmion radius $\lambda=10a$, the effective spin $\bar{S}=10$, $K=0.4J_1$, magnetic and electric fields $h\times 1$T, $E_z\times 250$V/m and dimensionless $E_z$ is varied. Triangular geometry enhances the spin frustration. In what follows the dimensionless Hamiltonian is obtained by dividing on the nearest exchange constant $H/J_1$, where $J_{1}=1$meV. $\hbar=1$ means that we are using dimensionless time unit $t\omega_0$ corresponding to $\omega_0=J_1/\hbar=10^2$GHz. Therefore, the theoretical approach developed in the present study also has material-specific aspects. 
\begin{figure}[t]
\centering
\includegraphics[width=0.45\textwidth]{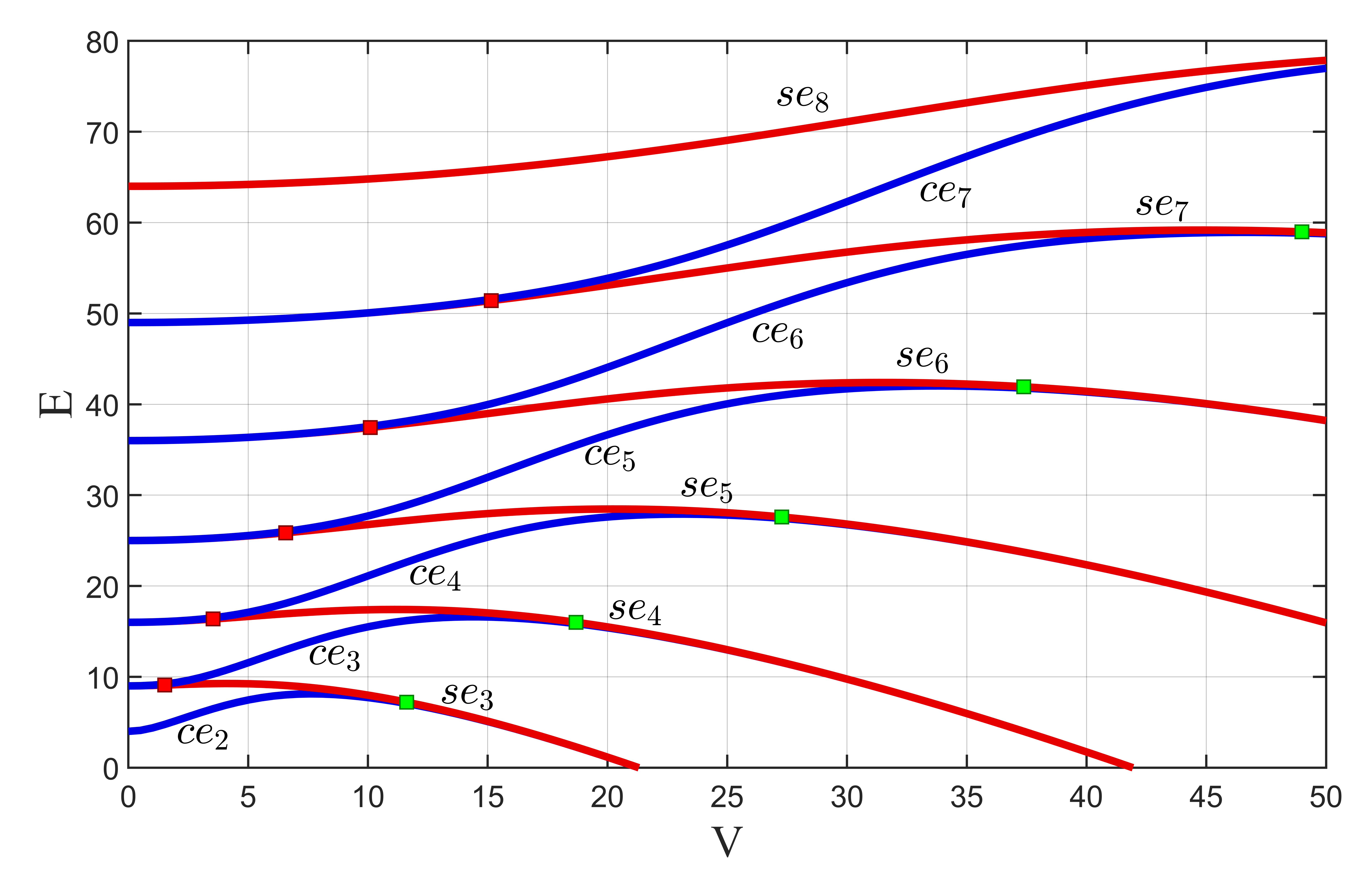}
\caption{Ince-Strutt Diagram: the energy spectrum of the Mathieu-Schr\"odinger equation as a function of the barrier height $E_n(V)$ obtained via numerical integration of Eq.(\ref{1Mathieu}). The parametric space contain the three subgroups of symmetry: $\mathcal{K}^{\pm}$ and $\mathcal{K}^{0}$. The border between the degenerated region $\mathcal{K}^{-}$ and $\mathcal{K}^{0}$ are given by red branching points whereas between $\mathcal{K}^{0}$ and right degenerated region $\mathcal{K}^{+}$ by green branching points. The size of those branching points is larger than $E_n/\bar{S}^3$, meaning that they cannot be found exactly ($E_n$ is the corresponding energy term and $\bar{S}$ is the effective spin). The eigenstates of $\mathcal{K}^{0}$ are given by Mathieu functions  whereas eigenstates of $\mathcal{K}^{\mp}$ by Eq.(\ref{irreducible representationK-}), Eq.(\ref{irreducible representationK+}) respectively. Starting with the initial state $\ket{\psi^+_{2n}}$ for small barrier $V(0)$ and increasing $V(t)$ in time we eventually populate four states $\ket{\phi^{\pm}_{2n}}$, $\ket{\phi^{\pm}_{2n-1}}$ forming the skyrmion qudit. }
\label{Ince-Strutt Diagram}
\end{figure}
 The quantum skyrmion model proposed in \cite{PhysRevLett.127.067201} is widely discussed in the literature but is solved only in the limit of a small barrier. In the present work, we derive a general analytic solution valid for an arbitrary barrier and demonstrate that the skyrmion is described by a quantum qudit state. We note that quantum qudits have particular advantages as compared to the qubits \cite{wang2020qudits,PRXQuantum.4.030327,nikolaeva2024efficient}. Namely, quantum qudits hold and save more quantum information than quantum qubits. For example, a qubit holds $\log_2(2)=1$ bit of quantum information, while a d-level Qudit $d>2$ holds $\log_2(d)$ bits of quantum information. Therefore, for saving the same amount of quantum information, one needs a much smaller number of qudits compared to qubits. One can therefore store and process quantum information with a smaller number of qudits. Besides, due to the smaller amount of information stored in individual qubits and the need for having more qubits, the architecture of qubit-based circuits becomes more complex and involved than it would be with qudits. Another critical question is the impact of hardware, environmental noise, and decoherence \cite{wang2020qudits,PRXQuantum.4.030327,nikolaeva2024efficient}. States of qudits are more robust and have better noise tolerance than qubit states. Qudit-based platform processing and holding the same amount of quantum information will be more compact. In what follows, we drive the quantum state via the external electric field $E_z$. Taking into account realistic values of the dimensionless anisotropy and Zeeman terms $k=K/J_1=0.1$, $h=g\mu_BB_z/J_1=0.47$, with the effective spin $\bar{S}=10$, we obtain an estimation of the strong electric field $E_z\approx 400V/m.$ For such values of the electric field we can reach the right (green) branch points in Fig.\ref{Ince-Strutt Diagram}, for quantum states up to the quantum number $n=7$.

\section{Model}

The authors of the work  \cite{PhysRevLett.127.067201}  exploited a method of collective coordinate quantization. In particular, authors performed canonical transformation in the phase space path integral and introduced canonical pair of variables \cite{PhysRevLett.127.067201}: $\hat\varphi_0$ and $\hat S_z$, with the commutation relations $[\hat\varphi_0, \hat S_z]=i/\bar{S}$. Here $\bar{S}$ is the effective spin that enters in the action $\mathcal{S}=\bar{S}\int dt\int dr[\dot{\Phi}(\Pi-1)-F(\Phi, \Pi)]$, the variable $\Pi=\cos\Theta (\textbf{r})$  is canonically conjugate to $\Phi(\textbf{r})$ and $S_z=\int\limits_\textbf{r}[1-\cos\Theta(\textbf{r})]\partial_\phi\Phi(\textbf{r})$. Two  angles $\Theta(\textbf{r})$, $\Phi(\textbf{r})$ characterize the two-dimensional magnetic texture with parametrization  $\textbf{m}(\textbf{r})=[\sin\Theta(\textbf{r})\cos\Phi(\textbf{r}), \sin\Theta(\textbf{r})\sin\Phi(\textbf{r}), \cos\Theta(\textbf{r})]$, $\textbf{r}=(x,y)$ and orthogonal symmetry $\textbf{m}(\textbf{r})\rightarrow\mathcal{\hat O}(2,\hat\varphi_0)\textbf{m}(\textbf{r})$. The rotation angle $\hat\varphi_0$ is the collective coordinate of the skyrmion helicity and $\hat S_z=(-i/\bar{S})\partial_{\hat\varphi_0}$ is its conjugate momentum: $\hat S_z\ket{s}=s/\bar{S}\ket{s}$, $\hat\varphi_0\ket{\varphi_0}=\varphi_0\ket{\varphi_0}$. The eigenstates are periodic functions $\ket{\varphi_0+2\pi}=\ket{\varphi_0}$.  Eventually, the problem of quantum skyrmion reduces to the effective Hamiltonian:
\begin{eqnarray}\label{initial}
\hat H_{S_z}=k(\hat S_z-h/k)^2-E_z\cos\hat\varphi_0,
\end{eqnarray}
where $k$ and $h$ characterize anisotropy and the external magnetic field, whereas $E_z$ is the applied external electric field. Considering rescaling $\varphi_0=2\varphi$, $V=-E_z4\bar{S}^2/k$, $\hat H_{S_z}=\hat H_{S_z}4\bar{S}^2/k$ we rewrite Hamiltonian Eq.(\ref{initial}) in the form
\begin{eqnarray}\label{transformed initial}
\hat H_{S_z}=-\frac{d^2}{d\varphi^2}+\frac{4ih\bar{S}}{k}\frac{d}{d\varphi}+\frac{4h^2\bar{S}^2}{k^2}+V\cos\varphi.
\end{eqnarray}
We note that the interaction of the
quantum skyrmion with the environment may lead to decoherence. The effect of the Ohmic damping terms for the operators $\hat\varphi_0$ and $\hat S_z$ was studied in \cite{PhysRevLett.127.067201} and it was found that the coherence time is quite large and is in the microsecond range. Therefore, we neglect the relaxation processes and consider the unitary dynamics.
In what follows, we assume that the electric field is adiabatically steered in time $V(t)=V_0+V_1t$ and $E$ is the characteristic energy scale of the Hamiltonian Eq.(\ref{transformed initial}). We set Planck's constant to $\hbar=1$. Then the Schr\"odinger equation we aim to solve converts into a non-stationary problem:
\begin{eqnarray}\label{schrodinger}
i\frac{d\ket{\Psi}}{dt}=\hat H_{S_z}[V(t)]\ket{\Psi}.
\end{eqnarray}
At first we neglect the time-dependent term in electric field and solve stationary problem. After transformation $\Phi=e^{i\bar{S}h\varphi/k}\bra{\varphi_0}\Psi\rangle$ we deduce the following Mathieu-Schr\"odinger equation \cite{chotorlishvili2010quantum,PhysRevE.71.056211}:
\begin{eqnarray}\label{1Mathieu}
\frac{d^2\Phi_n}{d\varphi^2}+(E_n-V_0\cos2\varphi)\Phi_n=0.
\end{eqnarray}
Mathieu-Schr\"odinger equation Eq.(\ref{1Mathieu}) has a discrete spectrum characterized by the non-trivial parametric dependence on the barrier height $V_0$. In the limit of small $V_0\rightarrow 0$, we recover
\cite{PhysRevLett.127.067201}, while the general case that we name Skyrmion Qudit is mathematically more demanding since the number of the levels exceeds two and depends on the modulation depth $V_1$ and the initial height of the barrier $V_0$. For the sake of brevity, $V_0$ and $V$ have the same meaning of barrier height if time-dependence $V(t)$ is not explicitly specified.

\section{Group theoretical analysis}

When deriving Eq.(\ref{initial}), the inverse effective spin $1/\bar{S}$ was considered as a small parameter \cite{PhysRevLett.127.067201} and higher order terms $\mathcal{O}(1/\bar{S}^3)$ were neglected. This approximation limits the accuracy of further discussion and imposes small uncertainty of the energy spectrum 
$\Delta E=E/\bar{S}^3$. The value of $\Delta E$ is decisive in pinpointing the split and merge bifurcation points of the energy spectrum. Those bifurcation points correspond to the critical values of the barrier height $V_{0n}(\mathcal{K}^-)$, $V_{0n}(\mathcal{K}^+)$ and define borders of the subgroups $\mathcal{K}^+$, $\mathcal{K}^-$, $\mathcal{K}^0$ of irreducible representation of the entire symmetry group of the problem $\mathcal{K}$. We proceed with the group theoretical analysis and explore the symmetry of the Mathieu-Schr\"odinger equation. The periodic solutions of Eq.(\ref{1Mathieu}) are given by the Mathieu functions \cite{bateman1953higher} for odd $m=2j+1$ and even $m=2j$, $j\in \mathbb{N}$ quantum numbers: $ce_{2m}(V,\varphi)$, $ce_{2m+1}(V,\varphi)$, $se_{2m+1}(V,\varphi)$, $s_{2m+2}(V,\varphi)$.  
The energy spectrum of the Mathieu-Schr\"odinger equation is described by the Mathieu characteristics plotted in Fig.\ref{Ince-Strutt Diagram}. The spectrum of the 
Mathieu-Schr\"odinger equation has nontrivial properties Fig.\ref{Ince-Strutt Diagram}. Namely for small barrier $V\in\mathcal{R}^-=[0, V_{0n}(\mathcal{K}^-))$ eigenvalues corresponding to the $se_n(V,\varphi)$ and $ce_n(V,\varphi)$ are merging $E_n^{se}(V)=E_n^{ce}(V)$. With the increase of $V$, degeneration in the spectrum is lifted at certain points $V_{0n}(\mathcal{K}^-)$ for each quantum level $n$. The bifurcation points $V_{0n}(\mathcal{K}^-)$ are shown in Fig.\ref{Ince-Strutt Diagram} by the red squares. With a further increase of the barrier $V\in\mathcal{R}^+=[V_{0n}(\mathcal{K}^+),\infty )$, energy levels merge, and degeneration occurs again. However, now the energy level $E_{2n}^{ce}(V)$ consolidates with $E_{2n+1}^{se}(V)$ and $E_{2n+1}^{ce}(V)$ with $E_{2n+2}^{se}(V)$, i.e., $E_{2n}^{ce}(V)=E_{2n+1}^{se}(V)$ and $E_{2n+1}^{ce}(V)=E_{2n+2}^{se}(V)$, shown by green squares in Fig.\ref{Ince-Strutt Diagram}. 
 We note again that model Eq.(\ref{1Mathieu}) is obtained in \cite{PhysRevLett.127.067201} via the expansion in terms of $1/\bar{S}$ (the effective spin $\bar{S}=10$). Since for triangular frustrated magnets the dominant terms are nearest and next nearest exchange interactions, the dipole-dipole interaction is also not taken into account. Small interactions not taken into account does not change the structure of energy levels but lead to their broadening. This effect is well-known in spectroscopy. The measured energy levels of complex quantum systems are not infinitely sharp but are characterized by a certain broadening \cite{sakurai2020modern}. Therefore, in dimensionless units $E/J_1$ ($J_1$ is the exchange constant), the resolution of energy terms is limited by $\Delta E\approx\mathcal{O}(1/\bar{S}^2)$. The broadening has particular consequences when we talk about the degeneracy of the spectrum. We can argue that the degeneracy in the spectrum is lifted only if the distance between two energy levels exceeds the level broadening $E(ce_n, V)-E(se_n, V)>\mathcal{O}(1/\bar{S}^2)$. We note also the quantum uncertainty. Two canonically conjugate variables in Hamiltonian do not commute $[\hat\varphi, \hat S_z]=i/\bar{S}$, meaning that we cannot measure simultaneously both variables. In the classical limit $\bar{S}\gg1$ variables commute, so that the Mathieu-Schrödinger equation converts to the classical Mathieu equation with well-known features. The spin and merging bifurcation points are defined via the accuracy $\Delta V_{0n}(\mathcal{K}^-)<V^-_{\text{max}}-V^-_{\text{min}}$ where maximal and minimal barrier height is defined from the equation $|E^{se}_n(V)-E^{ce}_n(V)|<\Delta E$, where $\Delta E=\text{sup}\lbrace E^{se}_n(V), E^{ce}_n(V)\rbrace/\bar{S}^2$. The same applies to the bifurcation points $\Delta V_{0n}(\mathcal{K}^+)<V^+_{\text{max}}-V^+_{\text{min}}$ where maximal and minimal barrier height is defined from the equation $|E^{se}_{2n+2}(V)-E^{ce}_{2n+1}(V)|<\Delta E$, where $\Delta E=\text{sup}\lbrace E^{se}_{2n+2}(V), E^{ce}_{2n+1}(V)\rbrace/\bar{S}^2$. Thus, we have three regions. Two of them, $\mathcal{R}^-$ and $\mathcal{R}^+$ for small and large $V$, respectively, correspond to the degenerate spectrum, while the region $\mathcal{R}^0$ laying in between $\mathcal{R}^-$ and $\mathcal{R}^+$ is not degenerate. The exact spectrum of the Mathieu-Schr\"odinger equation is obtained numerically and can be explained in terms of group-theoretical analysis. \vspace{0.2cm}\\
We consider the following symmetry operations $G(\varphi\rightarrow-\varphi)=\hat a$, $G(\varphi\rightarrow\pi-\varphi)=\hat b$, $G(\varphi\rightarrow\pi+\varphi)=\hat c$, $G(\varphi\rightarrow\varphi)=\hat e$. Mathieu functions under these operations transform as follows: $\hat e[ce_n(V,\varphi)]=ce_n(V,\varphi)$, 
$\hat e[se_n(V,\varphi)]=se_n(V,\varphi)$,
$\hat a[ce_n(V,\varphi)]=ce_n(V,\varphi)$, 
$\hat a[se_n(V,\varphi)]=-se_n(V,\varphi)$,
$\hat b[ce_{2n}(V,\varphi)]=ce_{2n}(V,\varphi)$, 
$\hat b[ce_{2n+1}(V,\varphi)]=-ce_{2n+1}(V,\varphi)$,
$\hat b[se_{2n+1}(V,\varphi)]=se_{2n+1}(V,\varphi)$, 
$\hat b[se_{2n+2}(V,\varphi)]=-se_{2n+2}(V,\varphi)$,
$\hat c[ce_{2n}(V,\varphi)]=ce_{2n}(V,\varphi)$, 
$\hat c[ce_{2n+1}(V,\varphi)]=-ce_{2n+1}(V,\varphi)$, and 
$\hat c[se_{2n+1}(V,\varphi)]=-se_{2n+1}(V,\varphi)$, 
$\hat c[se_{2n+2}(V,\varphi)]=se_{2n+2}(V,\varphi)$. With the associative binary
operation $G\times G\rightarrow G$ transformation operations form the Klein four-group $\mathcal{K} =\mathbb{Z}_2\times\mathbb{Z}_2$. Taking into account that 
$\hat e=\begin{pmatrix} 1 & 0\\0 & 1\end{pmatrix}$, $\hat a=\begin{pmatrix} 1 & 0\\0 & -1\end{pmatrix}$, $\hat b=\begin{pmatrix} -1 & 0\\0 & 1\end{pmatrix}$, $\hat c=\begin{pmatrix} -1 & 0\\0 & -1\end{pmatrix}$
and algebra $\hat a\star\hat a =\hat b\star\hat b=\hat c\star\hat c =\hat e$, $\hat a\star\hat b=\hat c$, $\hat a\star\hat c=\hat b$, $\hat b\star\hat c=\hat a$
we identify three normal subgroups:
\begin{eqnarray}\label{subgroups}
&&\mathcal{K}^+\Rightarrow\lbrace\hat e,\hat b\rbrace, \,\,\,\mathcal{K}^0\Rightarrow\lbrace\hat e,\hat c\rbrace,\\
&&\mathcal{K}^-\Rightarrow\lbrace\hat e,\hat a\rbrace.\nonumber
\end{eqnarray}
Within our interest are three asymptotic cases [\textbf{i}] small barrier limit $V\rightarrow 0$, [\textbf{ii}] moderate barrier limit $V\approx E$ (where $E$ is the expectation value of the energy, [\textbf{iii}] high barrier limit $V\gg E$. \vspace{0.2cm}\\
$[\textbf{i}]$. The  small barrier $V$ case, the region  $\mathcal{R}^-\in\mathcal{K}^-$.\vspace{0.2cm}\\ 
The existence of subgroups  $\mathcal{K}^+$, $\mathcal{K}^-$, $\mathcal{K}^0$, Eq.(\ref{subgroups}) hints on the degenerate eigenstates of the system with a higher symmetry rather than the symmetry defined by the entire group. The Abelian group $\mathcal{K}$ must have one-dimensional irreducible representations \cite{hassani2013mathematical}. The two-dimensional representation constructed in the base of real functions $se_n(V,\varphi)$ and $ce_n(V,\varphi)$ with degenerate spectrum  $E_{2n}^{ce}(V)=E_{2n+1}^{se}(V)$ in the limit of a small $V$ Fig.\ref{Ince-Strutt Diagram} is reducible and cannot serve as the true eigenstate. Note that true eigenstates of the degenerate spectrum should be complex
\cite{hamermesh2012group}. Thus irreducible representation of the subgroup $\mathcal{K}^-$ is given by $n\in\mathbb{N}$:
\begin{eqnarray}\label{irreducible representationK-}
\ket{\psi_n(V,\varphi)}=\frac{\sqrt{2}}{2}\left(\ket{ce_n(V, \varphi)}+i\ket{se_n(V, \varphi)}\right).
\end{eqnarray}
$[\textbf{ii}]$. The moderate barrier $V$ case, the region  $\mathcal{R}^0\in\mathcal{K}^0$.\vspace{0.2cm}\\The energy spectrum in the region $\mathcal{R}^0$ is non-degenerate and irreducible eigenstates
are $se_n(V,\varphi)$ and $ce_n(V,\varphi)$ for $n\in\mathbb{N}$. \vspace{0.2cm}\\
$[\textbf{iii}]$. The large barrier $V$ case, the region  $\mathcal{R}^+\in\mathcal{K}^+$.\vspace{0.2cm}\\ 
Similarly to the region $\mathcal{R}^-\in\mathcal{K}^-$, in the region $\mathcal{R}^+\in\mathcal{K}^+$, degenerated spectrum $E_{2n}^{ce}(V)=E_{2n+1}^{se}(V)$ and $E_{2n+1}^{ce}(V)=E_{2n+2}^{se}(V)$ is described by the irreducible eigenstates: 
\begin{eqnarray}\label{irreducible representationK+}
&&\ket{\phi_{2n}(V,\varphi)}=\ket{ce_{2n}(V,\varphi)}+i\ket{se_{2n+1}(V,\varphi)},\nonumber\\
&&\ket{\phi_{2n+1}(V,\varphi)}=\ket{ce_{2n+1}(V,\varphi)}+i\ket{se_{2n+2}(V,\varphi)}.
\end{eqnarray}
 
\section{Driving in time} 

It is a time to come back to the time-dependent problem. We assume that  
the skyrmion qudit is driven by a time-dependent electric field $V(t)=-\frac{4S^2}{k}(E_{0z}+E_1 t)=V_0+V_1(t)$. The modulation broadness is such that the system reaches four bifurcation points, as shown in Fig.\ref{Ince-Strutt Diagram}. Suppose that the system is initialized in the degenerate region $\mathcal{R}^-\in\mathcal{K}^-$ and the initial density matrix of the system reads $\hat\varrho[t=0, V_0]=\ket{\psi_{2n}(V_0,\varphi)}\bra{\psi_{2n}(V_0,\varphi)}$. To solve time-dependent problem for each region $\mathcal{R}^-$, $\mathcal{R}^0$, $\mathcal{R}^+$ we consider spectral decomposition of the evolution operator over the irreducible basis functions of the corresponding subgroups $\mathcal{K}^-$, $\mathcal{K}^0$, $\mathcal{K}^+$ respectively. The standard adiabatic approximation is formulated for non-degenerate spectrum only. However, Wilczek and Zee proposed extension of the adiabatic approximation for the degenerate spectrum  \cite{PhysRevLett.52.2111}. They showed that in case of degenerate spectrum, there appears an extra factor $\mathcal{P}\exp\left(-\int dt \langle\psi_n^+\ket{\partial_t{\psi}^-_n}\right)$, where $\mathcal{P}$ means time ordering. In our case Wilczek-Zee connection is zero $\langle\psi_n^+\ket{\partial_t{\psi}^-_n}=0$ and therefore we do not have an extra factor due to the degenerate spectrum. Taking into account the fact that driving is adiabatic. We exploit the adiabatic evolution operator proposed by M. Berry
\cite{berry2009transitionless}: 
\begin{eqnarray}\label{Berry1}
&&\hat U^{\mathcal{K}^-}(t_1)=\sum\limits_m\exp\left\lbrace-i\mathcal{D}^{\mathcal{K}^-}_m+\gamma^{\mathcal{K}^-}_m\right\rbrace\hat\Pi_m^{\mathcal{K}^-}.
\end{eqnarray}
Here, geometric  and dynamical phases are given by 
 $\gamma^{\mathcal{K}^-}_m=\int\limits_0^{t_1}dt'\bra{\psi_m(V(t'),\varphi)} \partial_{t'}\psi_m(V(t'),\varphi)\rangle$ and
$\mathcal{D}^{\mathcal{K}^-}_m=\int_0^{t_1}dt'E_m^{\mathcal{K}^-}[V(t')]$,  the operator defined in the subspace of irreducible representation of the subgroup $\mathcal{K}^-$ is given by $\hat\Pi_m^{\mathcal{K}^-}=\ket{\psi_m(V(t_1),\varphi)}\bra{\psi_m(V(0),\varphi)}$. 
Here $t_1$ is the time to reach the border between the regions $\mathcal{R}^-$ and $\mathcal{R}^0$, 
The evolved in time density matrix reads $ \hat\varrho[t_1, V(t_1)]=\hat U^{\mathcal{K}^-}(t_1)\hat\varrho[0, V_0]\hat (U^{\mathcal{K}^-}(t_1))^{-1}$ where $\hat\varrho[0, V_0]=\ket{\psi_{2n}(V_0,\varphi)}\bra{\psi_{2n}(V_0,\varphi)}$ is the initial state, $E_m^{\mathcal{K}^-}[V(t)]$ is the energy term in the region $\mathcal{R}^0$ and $\ket{\psi_m(V(t'),\varphi)}$ is the corresponding wave function of the irreducible representation of the subgroup $\mathcal{K}^-$. Due to the orthogonality of the state $\ket{\psi_n(V(t),\varphi)}$, from Eq.(\ref{Berry1}) we see that bra and ket vectors receive complex conjugate phases which cancel each other in the evolved density matrix and therefore $\hat\varrho[t_1, V(t_1)]=\ket{\psi_{2n}(V(t_1),\varphi)}\bra{\psi_{2n}(V(t_1),\varphi)}$. 
However, after reaching the bifurcation point $V(t)=V^{\mathcal{K}^-,\mathcal{K}^0}_c$, at the border between regions $\mathcal{R}^-$, $\mathcal{R}^0$, character of the time evolution changes. 
For calculation of the transition matrix elements, note that when the system reaches the left (red) branch point, according to Eq.(\ref{Berry1}) the evolved in time wave function reads 
\begin{eqnarray}\label{Berry1pluse}
&&\ket{\psi(t_1)}=\hat U^{\mathcal{K}^-}(t_1)\ket{\psi_{2n}(V(0),\varphi)}=\nonumber\\
&&\exp\lbrace-i\mathcal{D}^{\mathcal{K}^-}_{2n}+\gamma^{\mathcal{K}^-}_{2n}\rbrace\ket{\psi_{2n}(V(t_1),\varphi)}.
\end{eqnarray}
Due to the topology of energy levels Fig.\ref{Ince-Strutt Diagram}  upon reaching the red branching points there are two alternative transitions $\ket{\psi(t_1)}\rightarrow \ket{ce_{2n}(V, \varphi)}$ and $\ket{\psi(t_1)}\rightarrow \ket{se_{2n}(V, \varphi)}$. 
The transition probabilities  $\ket{\psi_{2n}(V,\varphi)}\rightarrow \ket{ce_{2n}(V, \varphi)}$ and $\ket{\psi_{2n}(V,\varphi)}\rightarrow \ket{se_{2n}(V, \varphi)}$ can be calculated directly according to the rules of quantum mechanics \cite{landau2013quantum}, i.e., $P_{\rightarrow ce_n}=\vert\langle\psi_{2n}(V,\varphi)\ket{ce_{2n}(V, \varphi)}\vert^2=1/2$ and similarly for $P_{\rightarrow se_n}=1/2$. It is easy to see that transition probabilities are equal to the expectation values of operators $\bra{\psi_{n}(V,\varphi)}\hat\Pi_{ce,n}^{\mathcal{K}^0}\ket{\psi_{n}(V,\varphi)}$, $\bra{\psi_{n}(V,\varphi)}\hat\Pi_{se,n}^{\mathcal{K}^0}\ket{\psi_{n}(V,\varphi)}$, where $\hat\Pi_{ce,m}^{\mathcal{K}^0}=\ket{ce_m(V(t_2))}\bra{ce_m(V(t_1))}$, 
$\hat\Pi_{se,m}^{\mathcal{K}^0}=\ket{se_m(V(t_2))}\bra{se_m(V(t_1))}$. Therefore, the evolution operator in the non-degenerate region takes the form:
\begin{eqnarray}\label{Berry2}
&&\hat U^{\mathcal{K}^0}(t_2)=\sum\limits_m\exp\left\lbrace-i\mathcal{D}^{\mathcal{K}^0}_{ce,m}+\gamma^{\mathcal{K}^0}_{ce,m}\right\rbrace\hat\Pi_{ce,m}^{\mathcal{K}^0}+\nonumber\\
&&\sum\limits_m\exp\left\lbrace-i\mathcal{D}^{\mathcal{K}^0}_{se,m}+\gamma^{\mathcal{K}^0}_{se,m}\right\rbrace\hat\Pi_{se,m}^{\mathcal{K}^0}.
\end{eqnarray}
The geometric phases are given by the expressions $\gamma^{\mathcal{K}^0}_{ce,m}=\int\limits_{t_1}^{t_2}dt'ce_m(V(t'),\varphi)\partial_{t'}ce_m(V(t'),\varphi)$ and $\gamma^{\mathcal{K}^0}_{se,m}=\int\limits_{t_1}^{t_2}dt'se_m(V(t'),\varphi)\partial_{t'}se_m(V(t'),\varphi)$, while dynamical phases $\mathcal{D}^{\mathcal{K}^0}_{ce,m}=\int_{t_1}^{t_2}dt'E_{ce,m}^{\mathcal{K}^-}[V(t')]$, $\mathcal{D}^{\mathcal{K}^0}_{se,m}=\int_{t_1}^{t_2}dt'E_{se,m}^{\mathcal{K}^-}[V(t')]$. Here in the region $\mathcal{R}^0$, for the spectral decomposition of the evolution operator, we used the basis functions of the irreducible representation of subgroup $\mathcal{K}^0$:  
$\hat\Pi_{ce,m}^{\mathcal{K}^0}=\ket{ce_m(V(t_2))}\bra{ce_m(V(t_1))}$, 
$\hat\Pi_{se,m}^{\mathcal{K}^0}=\ket{se_m(V(t_2))}\bra{se_m(V(t_1))}$. The time interval when the system is in the region $\mathcal{R}^0$ is given by  $t_2-t_1$.
 The density matrix of the skyrmion qudit after evolving in the region $\mathcal{R}^0$ reads:
\begin{eqnarray}\label{Berry3}
&&\hat\varrho^{\mathcal{K}^0}(t_2)=\frac{1}{2}\ket{ce_{2n}(V,\varphi)}\bra{ce_{2n}(V,\varphi)}+\nonumber\\
&&\frac{1}{2}\ket{se_{2n}(V,\varphi)}\bra{se_{2n}(V,\varphi)}-\nonumber\\
&&\frac{i}{2}e^{iF_{2n}}\ket{ce_{2n}(V,\varphi)}\bra{se_{2n}(V,\varphi)}+\nonumber\\
&&-\frac{i}{2}e^{-iF_{2n}}
\ket{se_{2n}(V,\varphi)}\bra{ce_{2n}(V,\varphi)}.
\end{eqnarray}
Here we introduce the notation for the sum of the accumulated dynamical and geometric phases 
$F_{2m}=[(\mathcal{D}^{\mathcal{K}^0}_{ce,2m}-\mathcal{D}^{\mathcal{K}^0}_{se,2m})+(\gamma^{\mathcal{K}^0}_{ce,{2m}}-\gamma^{\mathcal{K}^0}_{se,2m})]$ and omitted the time index $V(t_2)$. The same formalism can be applied to the transition between the regions $\mathcal{R}^0\rightarrow\mathcal{R}^+$. Due to the topology of the energy spectrum  we have the following alternative transitions
$\ket{ce_{2n}(V,\varphi)}\rightarrow \ket{\phi^{+}_{2n}(V, \varphi)}$,
$\ket{ce_{2n}(V,\varphi)}\rightarrow \ket{\phi^{-}_{2n}(V, \varphi)}$, 
$\ket{se_{2n}(V,\varphi)}\rightarrow \ket{\phi^{+}_{2n}(V, \varphi)}$,
$\ket{se_{2n}(V,\varphi)}\rightarrow \ket{\phi^{-}_{2n}(V, \varphi)}$.
Therefore the evolution operator after passing the border between subgroups $\mathcal{K}^0$, $\mathcal{K}^+$ takes the form:
\begin{eqnarray}\label{Berry4}
&&\hat U^{\mathcal{K}^+}(t>t_2)=\frac{1}{2}e^{g^+_{2n}(t)}\hat\Pi_{2n,+}^{\mathcal{K}^+}+\frac{1}{2}e^{g^+_{2n-1}(t)}\hat\Pi_{2n-1,+}^{\mathcal{K}^+}+\nonumber\\
&&\frac{1}{2}e^{g^+_{2n}(t)}\hat\Pi_{2n,-}^{\mathcal{K}^+}+\frac{1}{2}e^{g^+_{2n-1}(t)}\hat\Pi_{2n-1,-}^{\mathcal{K}^+}
\end{eqnarray}
Here we introduced expressions for the dynamical and geometric phases:
$g^+_{2n}(t)=-\mathcal{D}^{\mathcal{K}^+}_{2n,+}+\gamma^{\mathcal{K}^+}_{2n,+}$,
$g^-_{2n}(t)=-\mathcal{D}^{\mathcal{K}^+}_{2n,-}+\gamma^{\mathcal{K}^+}_{2n,-}$,
$g^+_{2n-1}(t)=-\mathcal{D}^{\mathcal{K}^+}_{2n-1,+}+\gamma^{\mathcal{K}^+}_{2n-1,+}$,
$g^-_{2n-1}(t)=-\mathcal{D}^{\mathcal{K}^+}_{2n-1,-}+\gamma^{\mathcal{K}^+}_{2n-1,-}$,
$\mathcal{D}^{\mathcal{K}^+}_{2n,+}=\mathcal{D}^{\mathcal{K}^+}_{2n,-}=\int_{t_2}^{t}dt'E_{2n}^{\mathcal{K}^+}[V(t)]$, $\mathcal{D}^{\mathcal{K}^+}_{2n-1,+}=\mathcal{D}^{\mathcal{K}^+}_{2n-1,-}=\int_{t_2}^{t}dt'E_{2n-1}^{\mathcal{K}^+}[V(t)]$, where:\\
$\gamma^{\mathcal{K}^+}_{2n,+}=\int\limits_{t_2}^tdt'\phi^+_{2n}(V(t'),\varphi)\partial_{t'}\phi^+_{2n}(V(t'),\varphi)$,\\
$\gamma^{\mathcal{K}^+}_{2n,-}=\int\limits_{t_2}^tdt'\phi^-_{2n}(V(t'),\varphi)\partial_{t'}\phi^-_{2n}(V(t'),\varphi)$,\\
$\gamma^{\mathcal{K}^+}_{2n-1,+}=\int\limits_{t_2}^tdt'\phi^+_{2n-1}(V(t'),\varphi)\partial_{t'}\phi^+_{2n-1}(V(t'),\varphi)$.\\
$\gamma^{\mathcal{K}^+}_{2n,-}=\int\limits_{t_2}^tdt'\phi^-_{2n-}(V(t'),\varphi)\partial_{t'}\phi^-_{2n-}(V(t'),\varphi)$. The four operators in Eq.(\ref{Berry4}) read:  
\begin{eqnarray}\label{Projectors}
&&\hat\Pi_{2n,+}^{\mathcal{K}^+}=\ket{\phi^+_{2n}(V(t),\varphi)}\bra{\phi^+_{2n}(V(t_2),\varphi)},\nonumber\\
&&\hat\Pi_{2n,-}^{\mathcal{K}^+}=\ket{\phi^-_{2n}(V(t),\varphi)}\bra{\phi^-_{2n}(V(t_2),\varphi)},\\
&&\hat\Pi_{2n-1,+}^{\mathcal{K}^+}=\ket{\phi^+_{2n-1}(V(t),\varphi)}\bra{\phi^+_{2n-1}(V(t_2),\varphi)},\nonumber\\
&&\hat\Pi_{2n-1,-}^{\mathcal{K}^+}=\ket{\phi^-_{2n-1}(V(t),\varphi)}\bra{\phi^-_{2n-1}(V(t_2),\varphi)}.\nonumber
\end{eqnarray}
For calculation of the time evolved state we utilized integrals from the Mathieu functions 
\cite{abramowitz1948handbook}. The skyrmion qudit state $\hat\varrho(t)=\hat U^{\mathcal{K}^+}(t>t_2)\hat\varrho^{\mathcal{K}^0}(t_2)(\hat U^{\mathcal{K}^+}(t>t_2))^{-1}$ after the evolution in the $\mathcal{R}^+$ region reads: 
\begin{eqnarray}\label{rhomatrix}
\hat\varrho(t)=\frac{1}{4}\begin{pmatrix}
1 & \rho_{12} & \rho_{13} & \rho_{14}\\
\rho^*_{12} & 1 & \rho_{23} & \rho_{24}\\
\rho^*_{13} & \rho^*_{24} & 1 & \rho_{34}\\
\rho^*_{14} & \rho^*_{24} & \rho^*_{34} & 1
\end{pmatrix},
\end{eqnarray}
where $\rho_{12}=-e^{i(g^-_{2n-1}-g^+_{2n-1})}$, $\rho_{13}=e^{i(g^-_{2n-1}-g^-_{2n}+F_{2n})}$, $\rho_{14}=e^{i(g^-_{2n-1}-g^+_{2n}+F_{2n})}$,
$\rho_{23}=-e^{i(g^+_{2n-1}-g^-_{2n}+F_{2n})}$, $\rho_{24}=-e^{i(g^+_{2n-1}-g^+_{2n}+F_{2n})}$, $\rho_{34}=e^{i(g^-_{2n}-g^+_{2n})}$.
The explicit form of the phases are presented in the appendix.
After performing POVM measurements based on the projectors Eq.(\ref{Projectors}) we derive the post-measurement state $\hat\varrho_{\text{post}}=\sum\limits_n\hat\Pi_n\hat\varrho(t)\hat\Pi_n/p_n$, $\hat\Pi_n=\ket{\phi_n(V,t)}\bra{\phi_n(V,t)}$, $p_n=\text{Tr}\lbrace\hat\Pi_n^\dag\hat\Pi_n\hat\varrho(t)\rbrace$:
\begin{eqnarray}\label{post-measurement state}
&&\hat\varrho_{\text{post}}=\frac{1}{4}\ket{\phi^+_{2n}(V(t),\varphi)}\bra{\phi^+_{2n}(V(t),\varphi)}+\nonumber\\
&&\frac{1}{4}\ket{\phi^-_{2n}(V(t),\varphi)}\bra{\phi^-_{2n}(V(t),\varphi)}+\\
&&\frac{1}{4}\ket{\phi^+_{2n-1}(V(t),\varphi)}\bra{\phi^+_{2n-1}(V(t),\varphi)}+\nonumber\\
&&\frac{1}{4}\ket{\phi^-_{2n-1}(V(t),\varphi)}\bra{\phi^-_{2n-1}(V(t),\varphi)}.\nonumber
\end{eqnarray}
The state Eq.(\ref{post-measurement state}) is the maximally mixed state with the entropy $S=-\log(1/4)$.
Calculation of the geometric phase in the general case is rather involved (see appendix). However, we can show that in the limit of small barrier the geometric phase is zero. Here we consider the particular case when the following equation holds:
\begin{eqnarray}\label{the following equation holds}
\frac{V^2(t)}{2(n^2-1)}<n^2, 
\end{eqnarray}
where $n$ is the quantum number and $V(t)$ is the barrier. Then Furrier expansion of the Mathieu functions simplifies: 
\begin{eqnarray}\label{ simplifies}
&&\psi_n(\varphi, V(t)) = \frac 1 {\sqrt{2}}
\left[ e^{in\varphi} - \frac{V}{4} \left( 
	\frac{e^{i(n-2)\varphi}}{n-1} - \frac{e^{i(n+2)\varphi}}{n+1} 
	\right) \right], \nonumber\\
&&\partial_t\psi_n(\varphi,V(t))=\frac{\partial\psi_n(\varphi,V(t))}{\partial V}\frac{dV(t)}{dt},
\end{eqnarray}
and for the geometric phase for  $V(t)=V_0t$ we deduce:
\begin{eqnarray}\label{twosimplifies}
&&\gamma^{\mathcal{K}^-}_m=
\int\limits_0^tdt\int\limits_{0}^{2\pi}d\varphi\Psi^*_n(\varphi,V(t'))\frac{\partial\Psi_n(\varphi,V(t)) }{\partial V}\frac{dV}{dt}.
\end{eqnarray}
We normalize wave function obtained via the perturbation theory:
\begin{equation}
	\label{eq:normalized_wave_function}
	\Psi_n(\varphi, V(t)) = \frac {\psi_n(\varphi, V(t))}{N},
\end{equation}
where
\begin{equation}
	\label{eq:normalization factor}
	N=\sqrt{\braket{\psi_n(\varphi, V(t)) | \psi_n(\varphi, V(t))}}.
\end{equation}
The normalization factor can be easily calculated
by taking integral over $\varphi$ and using identity
$\braket{\mathrm{exp}(-im\varphi)|\mathrm{exp}(n\varphi)} = 2\pi\delta_{nm}$:
\begin{equation}
	\label{eq:normalization_factor_final}
	N^2 = \pi\left[1+ \frac{V^2}{8(n^2-1)}
	\left(\frac{n^2+1}{n^2-1} - \delta_{n+2,n-2} \right)\right],
\end{equation}
It is easy to see that:
\begin{equation}
\label{eq:normv}
	\frac{\partial N}{\partial V} = \frac{1}{VN}(N^2 - \pi).
\end{equation}
and
\begin{equation}
\label{eq:derivative}
	\frac{\partial \psi_n(\varphi, V(t))}{\partial V} =
	\frac{1}{V} \left[\psi_n(\varphi,V(t)) - 
	\frac {1}{\sqrt{2}} e^{in\varphi}\right].
\end{equation}
Then, the value of geometric phase in the limit of small barrier can be calculated analytically and it is zero:
\begin{eqnarray}
	&&\gamma^{\mathcal{K}^-}_m=\int\limits_{0}^tdt\frac{1}{N^2}\int_0^{2\pi} d\varphi \psi_n^{*}(\varphi,V(t))\bigg[
	\frac{\partial\psi_n(\varphi, V(t))}{\partial V} -
\nonumber
\\
&&\psi_n(\varphi,V(t)) \frac {1}{N^2}(N^2-\pi)\bigg]. 
\end{eqnarray}
With further simplification 
\begin{eqnarray}
&&\gamma^{\mathcal{K}^-}_m=\frac {1}{N^2} \bigg[\int\limits_{0}^tdt\int_0^{2\pi} d\varphi \psi^{*}_n(\varphi,V(t))
	\frac{\partial \psi_n(\varphi,V(t))}{\partial V} -
	\nonumber
	\\
	&& \frac{1}{V}(N^2 - \pi)\bigg],
\end{eqnarray}
and finally
\begin{eqnarray}	
&&\gamma^{\mathcal{K}^-}_m=\frac{1}{N^2}\bigg[\frac{N^2}{V} 
	- \frac{1}{V\sqrt{2}}\int_0^{2\pi}d\varphi \psi^{*}(\varphi,V(t) )e^{in\varphi} -
	\nonumber
	\\
	&& \frac{1}{V}(N^2 - \pi) \bigg] = 0.
\end{eqnarray}

\section{Multivalued Logic and Coherence}
 
The skyrmion qudit is described by four quantum states:
\begin{eqnarray}\label{qudit states}
&&\ket{\phi^+_{2n}(V(t),\varphi)},\nonumber\\
&&\ket{\phi^-_{2n}(V(t),\varphi)},\nonumber\\
&&\ket{\phi^+_{2n-1}(V(t),\varphi)},\nonumber\\
&&\ket{\phi^-_{2n-1}(V(t),\varphi)}.
\end{eqnarray}
We note that the multivalued logic (MVL) is a type of logic system where variables can have more than just the two traditional values (0 and 1) of classical binary logic. Qudits have a natural ability to represent multivalued logic, and that is one of their advantages over qubits \cite{suleimenov2023improving}. The main advantage of the multi-valued logic is a larger coherence as compared
to the qubit states. We note that coherence in quantum information theory is viewed as a resource for performing quantum operations \cite{RevModPhys.89.041003}. The $l_1$ norm of coherence is given by the expression:
\begin{eqnarray}\label{coherence1}
C_{l_1}(\hat\varrho)=\sum\limits_{i\neq j}\vert\varrho_{ij}\vert.
\end{eqnarray}
At first we calculate coherence for the skyrmion cubit studied in \cite{PhysRevLett.127.067201}:
\begin{eqnarray}\label{coherence2}
\hat H_q=\frac{H_0}{2}\hat\sigma_z-\frac{X_c}{2}\hat\sigma_x.
\end{eqnarray}
Taking into account Eq.(\ref{coherence2}) for the coherence of skyrmion qubit we obtain:
\begin{eqnarray}\label{coherence3}
C^{qb}_{l_1}(\hat\varrho)=\frac{X_c\left(H_0/2+\frac{\sqrt{H_0^2+X_c^2}}{2}\right)}{\sqrt{H_0^2+X_c^2}+(X_c/2)^2}.
\end{eqnarray} 
We consider values of parameters used in \cite{PhysRevLett.127.067201}: $H_0=\kappa(1-2\bar{h})/\bar{S}$, $\bar{h}=h\bar{S}/\kappa$, $X_c=E_z$, with 
$\bar{S}=10$, $E_z=0.02$, $\kappa=0.1$, $h=0.47$ and obtain $C^{qb}_{l_1}(\hat\varrho)\approx 0.01$. As we see, the coherence of the skyrmion qubit is marginally small. The coherence of the skyrmion qudit can be calculated using Eq.(\ref{rhomatrix}): $C^{qD}_{l_1}(\hat\varrho)=12\gg C^{qb}_{l_1}(\hat\varrho)$.

\section{Conclusions}

Skyrmions in frustrated triangular magnets are formed due to the competing nearest ferromagnetic and next-nearest antiferromagnetic exchange interactions. They are characterized by a smaller radius $50\text{\AA}$ and described as quantum objects. Quantum skyrmions have highly unusual properties as compared to the classical skyrmions
and, due to their quantumness, cannot be described by continuous magnetic textures akin to the classical skyrmions. It is well known that skyrmions in frustrated magnets are characterized by the quantized helical degree of freedom and, therefore, can store quantum information. Nevertheless, before our work, the problem was solved only within the limits of the weak electric field, when the system can be described by a two-level model termed skyrmion qubit. In the present work, we studied quantum skyrmion and analyzed Skyrmion helicity and its conjugate momentum. We showed that a quantum skyrmion with quantized helicity is described by a skyrmion qudit model and obtained an exact analytic solution of the problem. We showed that a time-dependent Mathieu Schrödinger equation describes the system's evolution in time. Via the group theoretical analysis, we explored the symmetry properties of the Skyrmion Mathieu Schrödinger equation,  calculate level populations and transitions between them. The obtained results are of high interest for quantum information theory, quantum metrology, and skyrmionics. 
We note that proposed skyrmion qudit state has certain advantages as compared to the skyrmion qubit state. It can store more quantum information and is more stable to environmental effects and thermal noise.
We calculated the $l_1$ norm of coherence and showed that the coherence of the skyrmion quantum qudit is a thousand times larger than the coherence of the skyrmion quantum qubit. The obtained result is important for the perspectives of quantum skyrmion-based resource theory. 

\appendix
\section{Appendix}

The geometric phases:
\begin{eqnarray}\label{The geometric phases1}
&&\gamma^{\mathcal{K}^-}_m=\int\limits_0^{t_1}dt'\bra{\psi_m(V(t'),\varphi)} \partial_{t'}\psi_m(V(t'),\varphi)\rangle,\nonumber\\
&&\gamma^{\mathcal{K}^0}_{ce,m}=\int\limits_0^{t_1}dt'\bra{ce_m(V(t'),\varphi)} \partial_{t'}ce_m(V(t'),\varphi)\rangle,\nonumber\\
&&\gamma^{\mathcal{K}^0}_{se,m}=\int\limits_0^{t_1}dt'\bra{se_m(V(t'),\varphi)} \partial_{t'}se_m(V(t'),\varphi)\rangle,
\end{eqnarray}
and 
\begin{eqnarray}\label{The geometric phases2}
&&\gamma^{\mathcal{K}^+}_{2n,+}=\int\limits_0^{t_1}dt'\bra{\phi^+_{2m}(V(t'),\varphi)} \partial_{t'}\phi^+_{2m}(V(t'),\varphi)\rangle,\nonumber\\
&&\gamma^{\mathcal{K}^-}_{2m,-}=\int\limits_0^{t_1}dt'\bra{\phi^{-}_{2m}(V(t'),\varphi)} \partial_{t'}\phi^{-}_{2m}(V(t'),\varphi)\rangle,\nonumber\\
&&\gamma^{\mathcal{K}^+}_{2m-1,+}=\int\limits_0^{t_1}dt'\bra{\phi^+_{2m-1}(V(t'),\varphi)} \partial_{t'}\phi^+_{2m-1}(V(t'),\varphi)\rangle,\nonumber\\
&&\gamma^{\mathcal{K}^-}_{2m,-}=\int\limits_0^{t_1}dt'\bra{\phi^-_{2m-1}(V(t'),\varphi)} \partial_{t'}\phi^-_{2m-1}(V(t'),\varphi)\rangle.\nonumber
\end{eqnarray}
To calculate geometric phases we take into account that $\ket{\phi_{2m}(V,\varphi)}=\ket{ce_{2m+1}}+i\ket{se_{2m+1}}$, $\ket{\phi_{2m+1}(V,\varphi)}=\ket{ce_{2m}}+i\ket{se_{2m+2}}$, 
$\ket{\psi_{2m}(V,\varphi)}=\ket{ce_{2m}}+i\ket{se_{2m}}$,  and trigonometric representations
$\ket{ce_{2m}}=\sum\limits_{r=0}^\infty A_{2r}^{2m}(V)\cos2r\varphi$, 
$\ket{se_{2m}}=\sum\limits_{r=0}^\infty B_{2r}^{2m}(V)\sin2r\varphi$, 
$\ket{ce_{2m+1}}=\sum\limits_{r=0}^\infty A_{2r+1}^{2m+1}(V)\cos(2r+1)\varphi$,
$\ket{se_{2m+1}}=\sum\limits_{r=0}^\infty B_{2r+1}^{2m+1}(V)\sin(2r+1)\varphi$.
For illustration we explicitly calculate $\gamma^{\mathcal{K}^-}_m$, $m=2n+1$: 
To calculate integral we exploit time-dependent Mathieu-Schr\"odinger equation and rewrite integral in equivalent form:
\begin{eqnarray}
&&\gamma^{\mathcal{K}^-}_m=i\int\limits_0^{t_1}dt'\bra{\psi_m}\partial_{t'}\psi_m\rangle=\nonumber\\
&&\int\limits_0^{t_1}dt'\bra{\psi_m}\hat H(t')\ket{\psi_m}.
\end{eqnarray}
Then using Furrier expansion of Mathieu functions we deduce:
\begin{eqnarray}\label{we explicitly calculate}
&&\gamma^{\mathcal{K}^-}_m=(1/4)\int\limits_0^{t_1}V(t)\sum\limits_{r=0}^\infty A_{2r+1}^m(V(t))A_{2r+3}^m(V(t))dt+\nonumber\\
&&(1/4)\int\limits_0^{t_1} V(t)\sum\limits_{r=0}^\infty A_{2r+1}^m(V(t))A_{2r-1}^m(V(t))dt+\nonumber\\
&&(1/4)\int\limits_0^{t_1}V(t)\sum\limits_{r=0}^\infty B_{2r+1}^m(V(t))B_{2r+3}^m(V(t))dt+\\
&&(1/4)\int\limits_0^{t_1}V(t)\sum\limits_{r=0}^\infty B_{2r+1}^m(V(t))B_{2r-1}^m(V(t))dt+\nonumber\\
&&+(1/8)\int\limits_0^{t_1}\sum\limits_{r=0}^\infty V(t)(2r+1)^4[A_{2r+1}^m(V(t))]^2[B_{2r+1}^m(V(t))]^2.\nonumber
\end{eqnarray}

\bibliography{Dimitris1}

\end{document}